\documentstyle[12pt]{article}

\begin{document}
\begin{titlepage}
\thispagestyle{empty}
\title{ Remarks on the Calculation of the Enhancement Factor in the $K \rightarrow 2\pi$
Decay }
\vspace{4.0cm}
\author{Tran N. Truong \\
\small \em Centre de Physique Th{\'e}orique, 
{\footnote {unit{\'e} propre 014 du
CNRS}}\\ 
\small \em Ecole Polytechnique \\
\small \em F91128 Palaiseau, France}

\date{August 2000}

\maketitle

\begin{abstract}
The problem of  calculation of the enhancement factor due to the strong pion final state
interaction is reexamined in the light of  recent interest in  understanding of the
origine of the $\Delta I=1/2$ rule and calculations of the CP violation parameter
$\epsilon^{\prime}/\epsilon$. It is shown that the traditional method of calculating
the \emph{absolute} enhancement factor is model dependent, while the method of relating $K
\rightarrow 2\pi$ amplitude to the  $K-\pi$ matrix element using Current Algebra is on a 
safer ground.

\end{abstract}
\end{titlepage}

The problem of calculating the absolute enhancement factor due to the final state
interaction is a well-known problem in particle physics \cite{isgur, brown}. The solution for
the hadronic final state interaction problem, if it existed at all, could be quite useful in
understanding many problems in the low energy particle physics, in particular the origine of
the $\Delta I=1/2$ rule in the Kaon decay and  the  ratio
$\epsilon^{\prime}/\epsilon$  in the CP violation of the
$K 
\rightarrow 2\pi$ decays \cite{truong1}-\cite{truong3}.

 There are two approaches in the litterature. One is the classical 
potential  method, using  the well-known result of the  correction due to the
Coulombian final state interaction  in  
$\beta$ decays \cite{isgur, brown}. In this approach, the interacting  wave
function is calculated at the origin  due to the relatively 
  long range strong interaction of the two hadrons in the final state  compared with the
shorter range production or decay mechanisms.  This method is called, in the following, the
potential approach. Here the \emph{absolute} enhancement factor is calculated.

 Another  approach \cite{truong1}-\cite{truong3} to this problem is
not to calculate the absolute enhancement factor, but only to calculate the energy variation
of the enhancement factor between two energy scales which are not far apart. Using 
current algebra low energy theorem \cite{adler} or the effective Lagrangian, one process can
be related to another. 

For example, the 
$K \rightarrow 2\pi$, $K
\rightarrow 3\pi$ are related to the $K- 
\pi$  mixing. Similarly the $\eta \rightarrow 3\pi$ rate is
related to the
$\eta-
\pi$ mixing; the $K \rightarrow \pi\pi e\nu$ ($K_{l4}$) is related to $K
\rightarrow
\pi e\nu$ ($K_{l3}$) and $K\rightarrow e\nu$ ($K_{l2}$) amplitudes. In our problem, the
$K\rightarrow 2\pi$ is related to the $K-\pi$ mixing. The latter process, having no final
state interaction, can be calculated by the numerical method of the lattice gauge theory or
by some approximation schemes. 
In the following, for
simplicity, this method is called as the current algebra approach.

The purpose of this note is to compare these two approaches for $K \rightarrow 2\pi$ decay.
It is pointed out that these apparently unrelated approaches are in fact similar, 
but the potential approach is unreliable while the current algebra is more believable. They 
both describe  the  energy variation of the Omnes function \cite {omnes} or enhancement
factor at two energy points. In the potential approach, it is the variation of the
(unsubtracted) Omnes function  at infinite energy, compared with the Omnes function
evaluated at the Kaon mass is calculated. In the current algebra approach, the variation of
the enhancement factor between two points, the soft  current algebra  and the  physical
matrix element points, is calculated;  they are relatively near which makes this type of
calculation reliable.

\section{Absolute Enhancement Factor in $K \rightarrow 2\pi$ Decay}

Let us first examine the potential approach. The effect of the final state interaction on
the matrix element is just the wave function of the interacting final state hadrons
evaluated at the origin.

We first want to show that the wave function at the origin does indeed have the strong
interacting phase, it has the \emph{ unsubtracted} Omnes form, and then we want to show that 
its magnitude is, however, model dependent.

Under some restrictions on the conditions  on the integral of the interacting potential
$V(r)$ as a function of the distance, namely:
\begin{equation}
 \int_{0}^\infty dr r^m V(r) < \infty
\end{equation}
where $m=1$ and $m=2$. These are the conditions on the short range behavior of the
potential.  Under these restrictions, some rigourous results are known on the wave function
at the origin.

The wave function at the origin or the enhancement factor is given by the inverse of the
Jost function
$f(-k)$, where k is the the relative momentum of the two-body system \cite{jost}. In the
energy plane,
$s=4(m_\pi^2+k^2)$, the inverse of the Jost function is analytic in the cut plane with a
branch point at
$s=4m_\pi^2$; the zeros of the Jost function are on the real $s$ axis and correspond to the
bound states. The phase of the inverse of the Jost function, $1/f(-k)$, is the two-body
elastic phase shifts $\delta$. For $s\rightarrow\infty$, the inverse of the Jost function has
a well-defined limit: it is real and equal to unity.( For a review of the potential theory
see
\cite{goldberger, gillespie, treiman}). Because of these conditions, the inverse of the
Jost function, which is denoted in the following as $1/D(s)$, has the following integral
representation (assuming that there are no bound states):
\begin{equation}
\frac{1}{D(s)}=\exp(\frac{1}{\pi} \int_{4m_\pi^2}^\infty \frac{\delta(z)dz}{z-s-i\epsilon})
\label{eq:jost}
\end{equation} 
This equation can be derived by applying the Cauchy theorem for $\ln(1/D(s))$, using the fact
that there are no bound states and hence the Jost function $f(-k)$ has no zeroes in the $s$ 
plane and is equal to unity at infinity.  The convergence of the integral
in Eq. (\ref{eq:jost}) is assured by the restriction of the potential which demands that the
phase shifts tend to  zero sufficiently fast as $s\rightarrow \infty$.  At
an energy
$s$, the enhancement factor is just given by Eq. (\ref{eq:jost}). The physical value of the
matrix element is obtained by setting $s=m_K^2$ where $m_K$ is the Kaon mass. (If there were
bound states, the expression for the wave function would be slightly more  complicated and
the variation of the phase shifts between zero and infinite energy must obey the Levinson's
theorem
\cite{levinson},
$\delta(0)-\delta(\infty)$=(number of bound states)$\pi$, with
$\delta(\infty)=0$). Because there are no bound states in $\pi\pi$ scattering, the S-wave
$I=0$ phase shift at the $2\pi$ threshold is zero and has to go to zero at $s=\infty$
sufficiently fast.

From these considerations, a calculation of the absolute
enhancement factor implies implicitly  a variation of the $1/D(s)$ factor between $s=\infty$
where it is unity, and
$s=m_K^2$. This is a very large energy region to calculate the
 variation of the enhancement factor, and hence it is difficult to get a reliable result.

This sensitivity can also be seen from the study of the scattering of two particles by a
potential. Although this problem was studied by the authors of the reference \cite{isgur} who
claimed that the enhancement factor is insensitive to the constructed potentials, we wish
to point out that the two potentials that they constructed are quite similar, one is the
square well and the other is Gaussian.  Because they
are quite similar in both the short and long range behaviors, one cannot claim the
insensitivity of the calculation of the enhancement factor.

In fact, it is well-known
that the value of the wave function at the origin depends strongly on the  behavior of the
potential near the origin but the low energy behavior of the phase shifts does not.
 One can easily construct a
potential with an inner repulsive core but  with an outer attractive part to give
approximately the same scattering length and effective range as those given by a purely
attractive potential. Yet the wave functions at the origin are quite different for both
cases. In the purely attractive potential the modulus of the enhancement factor is larger
than unity. In the case of a more complicated potential, with an inner repulsive core and
an outer attractive part, the modulus of the enhancement factor could be less than
 unity. In fact with an infinite repulsive core, the wave function at
the origin  is zero.

This result can also be seen from 
 the integral representation of the inverse of the Jost function in terms of the
elastic phase shifts, i.e from the \emph{unsubtracted} form of the Omnes function, Eq.
(\ref{eq:jost}). It is clear that the enhancement factor evaluated at $s=m_K^2$ is sensitive
to the asymptotic behavior of the phase shifts corresponding to the shorter range of the
potential.  A subtracted form of the Omnes function is less sensitive to the high energy
behavior of the phase shifts due to the weight factor in the integral representation (see
below).

 Although
some qualitative features of the potential approach may be correct, it is difficult to make
the calculation scheme reliable because one has effectively to calculate the variation of
the  enhancement factor at
$s=m_K^2$  compared to that at infinite energy (which is unity). Our viewpoint is
therefore the unsubtracted form of the Omnes function or the absolute enhancement factor,
being sensitive to the high energy behaviour of the phase shifts, should be avoided in
theoretical calculations. It can however be used with confidence to study the energy
dependence of the matrix element as will be shown below.

 \section{Current Algebra Low Energy Theorem: Relation between $K \rightarrow 2\pi$ and the
off-shell
$K-\pi$ transition}
The above approach is in contrast with the calculation of the variation of the
function 
$1/D(s)$ at low energy. This is a typical problem one
has to deal with when Chiral Symmetry is relevant. To make this point clear, let us consider
the chain $K_{l4}, K_{l3}, K_{l2}$. They are related to each other by the soft current
algebra theorems and also by the Effective Lagrangian. Roughly speaking, in the limit  of one
the pion soft in the
$K_{l4}$ decay, its matrix element is equal to the $K_{l3}$. This relation is independent of
whether there is a strong final state interaction between the two outgoing pions or not.

 Using this idea, the problem of relating the $K \rightarrow 2\pi$ and the
off-shell
$K-\pi$ transition, was examined a long time ago \cite{truong1,
truong2}. There are recently questions how
various formula in ref. \cite{truong1, truong2} were obtained \cite{ neubert, buras}. A more
detailed explanation can be found in  \cite{truong3}. We briefly summarized here how these
results were obtained. The effective Lagrangian for the
$K\rightarrow 2\pi$ is given by: 
\begin{equation}
M(K_S(k)\rightarrow \pi^+(p)+\pi^-(q) )= \frac{i}{\sqrt{2}}Cf_\pi(2k^2-p^2-q^2)
\label{eq:k2pi}
\end{equation}
and
\begin{equation}
 M(K_L\rightarrow\pi^0)=-C\sqrt{2}f_\pi^2q(\pi).q(K)
\label{eq:kpi}
\end{equation}
Hence Eqs. ({\ref{eq:k2pi}, \ref{eq:kpi}) are related to each other by a Clebsch-Gordan
coefficients and a factor of $f_\pi$ in the limit of $p_\mu \rightarrow 0$. This is
typically a current algebra low energy theorem. Unlike other current algebra results, the
matrix elements given here are strongly energy dependent.

Let us  consider the matrix element
of
$K(k)\rightarrow
\pi^+(p)+\pi^-(q)$ as function of the complex variables $k^2=s$ with first the two pion on
their mass shell. This matrix element is an analytic function in $s$ with a cut starting from
$4m_\pi^2$ to $\infty$. The imaginary part of this matrix element can have contribution from
the self energy graphs and also from the contribution from the unitarity relation
$2\pi,4\pi$,... intermediate states. Because it is an analytic function with a cut on the
real axis from
$4m_\pi^2$ to $\infty$, this matrix element  can be expanded as a power
series in
$s-s_0$ where
$s_0$ is outside the cut.  Eq.
(\ref{eq:k2pi}) should be interpreted in this way and therefore includes the final state
interaction effect. 

Furthermore the Cabibbo-Gell-Mann Theorem \cite{cab} requires that the $K\rightarrow
2\pi$ matrix element has to vanish in the  $SU(3)$ limit which is clearly satisfied by Eq.
(\ref{eq:k2pi}). Using this condition, one has to take the expansion point at $s_0=m_\pi^2$
\cite{truong1, truong2, truong3}.

The $K \rightarrow 2\pi$ matrix element is therefore (with $k^2=s$ and the pions are on
their mass shell):
\begin{equation}
M(K_S(s)\rightarrow \pi^+\pi^- )= \sqrt{2}Cf_\pi(s-m_\pi^2) \frac{1}{D(s,m_\pi^2)}
\label{eq:1}
\end{equation}
while the $K-\pi$ matrix element, with the pion on its mass shell, the Kaon off its mass
shell and the weak hamiltonian carries no momentum, is:
\begin{equation}
 M(K_L\rightarrow\pi^0)=-C\sqrt{2}f_\pi^2 m_\pi^2 \label{eq:2}
\end{equation}
where 
\begin{equation}
\frac{1}{D(s,m_\pi^2)} = \exp(\frac{s-m_\pi^2}{\pi} \int_{4m_\pi^2}^\infty
\frac{\delta(z)dz}{(z-m_\pi^2)(z-s-i\epsilon)}) \label{eq:1/d}
\end{equation}

Eqs. (\ref{eq:1}, \ref{eq:2}) give a relation between the $K-\pi$  and the $K \rightarrow
2\pi$ transitions. Physical value of $K \rightarrow 2\pi$ is obtained by setting $s=m_K^2$.

Because of the subtracted form of the Omnes function in Eq. (\ref{eq:1}), defined in Eq.
(\ref{eq:1/d}), the enhancement factor is not sensitive to the high energy behavior of the
phase shift (or to the inner part of the S-wave $\pi\pi$ potential). Here one studies the
variation of the enhancement factor between two near-by points, $s=m_\pi^2$ and $s=m_K^2$.

\section{Analogy with the Pion Form Factor Calculation}
In the current algebra method, two apparently unrelated processes are related to each other
by the current algebra soft pion theorem. Here the partially conservation of the axial
current and the current commutation relations play an important role. Effective Lagrangian
synthesizes these results in a simple manner.

It might be useful to compare the current algebra $K \rightarrow
2\pi$ problem with the corresponding vector pion form factor V(s) calculation. 

  The  conservation of the hadronic vector
current, together with the usual commutation relations in field theory, enables us to
 demonstrate the Ward identity at zero momentum transfer for the vector pion form factor
 $V(0)=1$ \cite{weinberg}. This condition sets the scale for the form factor calculation. The
calculation of the vector pion form factor is therefore similar to the relation between
$K\rightarrow 2\pi$ and
$K-\pi$ mixing. It is  of interest to see how the $\pi\pi$ final state interaction
effect due to the potential approach would give.

  Because the two pions are in the relative P-state, the first derivative of the
interacting wave function at the origin is relevant in calculating the final state
interaction effect. The enhancement factor is given by the inverse of the  P-wave Jost
function. The low energy theorem for V(0) as required by the Ward identity is violated
because the inverse of the Jost function at $s=0$ is no longer equal to unity. It is however
equal to unity at $s=\infty$ which is not required by any other general principle.

The potential approach to this problem has to be modified: one simply has to forget the
condition at infinite energy, but taking into account of the Ward identity 
condition at zero momentum transfer:
\begin{equation}
V(s)=D_1(0)/D_1(s) \label{eq:ff}
\end{equation}
where $D_1(s)$ is similar to Eq. (\ref{eq:jost}) with the S-wave phase shifts replaced by
the P-wave phase shifts. Eq. (\ref{eq:ff}) is the standard formula for the vector pion form
factor
$V(s)$.

\section{Conclusion}
 In conclusion, the absolute enhancement factor for $K\rightarrow 2\pi$ is model dependent,
but the relation between
$K\rightarrow 2\pi$ and $K-\pi$ using analyticity, unitarity and the effective Lagrangian or
current algebra,
 is more reliable.


\newpage
\begin{thebibliography}{99}
\bibitem{isgur} N. Isgur, K. Maltman, J. Weinstein and T. Barnes, Phys. Rev. Lett. {\bf 64}
(1990) 161.
\bibitem{brown} G. E. Brown, J. W. Durso, M. B. Johnson and J. Speth, Phys. Lett. {\bf 238}
(1990) 20.
\bibitem{truong1} T. N. Truong, Phys. Lett. {\bf B207} (1988) 495.
\bibitem{truong2} T. N. Truong, Phys. Lett. {\bf B313} (1993)  221.
\bibitem{neubert} M. Neubert and B. Stech, Phys. Rev. {\bf D44} (1991) 775.
\bibitem{pich} E. Pallante and A.  Pich,  Phys. Rev. Lett.{\bf 84} (2000) 2568.
\bibitem{buras} A. J. Buras, M. Ciuchini, E. Franco, G. Isidori, G. Martinelli and L. Silvestrini,
Phys. Lett. {\bf B480} (2000) 80.
\bibitem{truong3} T. N. Truong (ph-hep/0004185).
\bibitem{adler} For a review, see \emph{Current Algebra} by S. L. Adler and R. Dashen, W.A.
Benjamin, New York 1968.
\bibitem{omnes} N. I. Muskhelishvili, \emph{ Singular Integral Equations} (Noordhoof,
Groningen, 1953).  R. Omnes, Nuovo Cimento {\bf 8}  (1958) 316.
\bibitem{jost} R. Jost, Helv. Phys. Acta. {\bf22} (1947) 256.
\bibitem{goldberger} M. L. Goldberger and K. M. Watson, \emph{Collision Theory}, John Willey
and Sons, Inc., New York 1964.
\bibitem{gillespie} J. Gillespie, \emph{ Final State Interactions}, Holden-Day Inc., San
Francisco 1964.
\bibitem{treiman} R.Blankenbecler, M. L. Goldberger, N. Khuri and S. B. Treiman, Ann. of
Phys. {\bf 10} (1960) 62.
\bibitem{levinson} N. Levinson, Danske Videnskab. Selskab, Mat.-fys. Medd. {\bf 25}, No.9
(1949).
\bibitem{cab} N. Cabibbo, Phys. Rev. Lett. {\bf12} (1964)  62; M. Gell-Mann, Phys. Rev. Lett.
{\bf 12} (1964) 155. See also more recent references cited in \cite{buras}.
\bibitem{weinberg} S. Weinberg, \emph{Lectures on Elementary Particles and Quantum Field
Theory, 1970 Brandeis University Summer Institute in Theoretical Physics}, MIT Press,
Cambridge 1970.
 
\end{thebibliography}
\end{document}